\documentclass[11pt]{article}
\usepackage{amsmath}

\usepackage{relsize}
\usepackage{color}
\title{{\bf The electromagnetic self-force of a uniformly charged spherical ball}}
\author{ G. Vaman\\Institute of Atomic Physics, P. O. Box MG-6, Bucharest, Romania\\ (vaman@ifin.nipne.ro)}
\begin{document}
\maketitle
\begin{abstract}
\noindent
We calculate the electromagnetic self-force of a uniformly charged spherical ball moving on a rectilinear trajectory, neglecting the Lorentz contraction.
\end{abstract}
\section*{1. Introduction}
 In this paper, we calculate the electromagnetic self-force of a uniformly charged spherical ball, moving on a rectilinear trajectory, with a time-dependent
 velocity. To this end, we use the {\it ansatz} from \cite{h}, Eq. (4)
\[ \vec{F}_s=- \frac{d \vec{G}_s(t)}{dt}, \]
where $\vec{F}_s$ is the self-force and $\vec{G}_s(t)=(1/4\pi c)\int d \vec{r}\; \vec{E}(\vec{r},t) \times \vec{B}(\vec{r},t) $ is the self-field momentum of the charge distribution. \\
The charge density of a uniformly charged ball of total charge $e$ and radius $a$, in the proper frame, is
\[
\rho(\vec{r})=\frac{3e}{4\pi a^3}\theta(a-r),
\]
where $\theta$ is the Heaviside function.
We start our calculation from its multipole expansion
\begin{align}\label{0}
\rho(\vec{r})=3e\sum_{n=0}^{\infty} \frac{a^{2n}}{2^n n! (2n+3)!!} \Delta^n \delta(\vec{r}),
\end{align}
where $\delta$ is the Dirac delta function.
Eq. (\ref{0}) can be justified, for example, as in \cite{row}. We consider that the charge distribution is spherical in the laboratory frame, so our results are valid for small velocities. To obtain the charge density in the laboratory frame, we replace in the right hand side of (\ref{0}) $\vec{r}$ with $\vec{r}-\vec{i} w(t)$, where we supposed that the ball is moving along the $x-$axis, with the velocity $\dot{w}(t)$.

\section*{2.The calculation of the self-force}
Starting from Eq. (\ref{0}) we can calculate easily the electromagnetic potentials
\[ \phi( \vec{r}, t)= \int d\;  \vec{r}\ ' \; \frac{ \rho( \vec{r}\ ', t- |\vec{r}-  \vec{r}\ '|/c)}{ |\vec{r}-  \vec{r}\ '|}, \; \; \; \vec{A}( \vec{r}, t)= \frac{1}{c} \int d\;  \vec{r}\ ' \; \frac{ \vec{j}(  \vec{r}\ ', t- |\vec{r}-  \vec{r}\ '|/c)}{ |\vec{r}-  \vec{r}\ '|}, \]
and the fields
\[ \vec{E}(\vec{r},t)= - \nabla \phi(\vec{r},t) - \frac{1}{c}\frac{\partial}{\partial t} \vec{A}(\vec{r},t), \; \; \vec{B}(\vec{r},t)= \nabla \times \vec{A}(\vec{r},t). \]
We write bellow only the $y-$ and $z-$ components of the fields.
\begin{align}\label{1}
E_{y,z} (\vec{r},t)= 3e \sum_{n=0}^{\infty} \sum_{p=0}^{\infty}\frac{a^{2n} (-1)^{p+1}}{2^n n! (2n+3)!! p!} \Delta^n \partial_{y,z} \partial_x^p \frac{w^p\left( t- \frac{r}{c} \right)}{r},
\end{align}
\begin{align}\label{2}
B_{y,z}  (\vec{r},t) = \frac{3e}{c}\sum_{n=0}^{\infty} \sum_{p=0}^{\infty} \frac{a^{2n} (-1)^{p}}{2^n n! (2n+3)!! p!} \Delta^n \tilde{\partial}_{z,y} \partial_x^p \frac{\dot{w}\left(t- \frac{r}{c}\right) w^p\left( t-\frac{r}{c} \right)}{r},
\end{align}
where $\tilde{\partial_z}=\partial_z$ and $\tilde{\partial_y}=-\partial_y$.
The Fourier transforms of the above fields are
\begin{align}\label{3}
E_{y,z}(\vec{k},t)= 12 \pi e \sum_{p=0}^{\infty} \frac{i^{p+1} k_{y,z} k_x^p}{k p!} \frac{j_1(ka)}{ka} \int_0^{\infty} dr \sin(kr) w^p\left(t-\frac{r}{c}\right),
\end{align}
\begin{align}\label{4}
B_{y,z} (-\vec{k},t)= \frac{12 \pi e }{c}  \sum_{p=0}^{\infty} \frac{(-1)^p i^{p+1}\tilde{k}_{z,y} k_x^p}{kp!} \frac{j_1(ka)}{ka} \int_0^{\infty} dr \sin(kr) \dot{w} \left( t-\frac{r}{c}\right) w^p\left(t-\frac{r}{c}\right),
\end{align}
where $\tilde{k}_z=k_z$ and $\tilde{k}_y=-k_y$, $j_1(x) = \sin(x)/x^2- \cos(x)/x$ is the spherical Bessel function of the first kind, and the Fourier transform of a function $f(\vec{r})$ is defined as $f(\vec{k})=\int d\vec{r}\; e^{i\vec{k} \vec{r}} f(\vec{r})$. To obtain Eqs. (\ref{3}), (\ref{4}), we used
\[ \sum_{k=0}^{\infty} \frac{(-1)^k x^{2k}}{2^k k! (2k+3)!!}= \frac{j_1(x)}{x}. \]
We calculate the self-force by integrating in the Fourier space
\begin{align}\label{5}
F_x= & - \frac{1}{4\pi c} \frac{d}{dt} \int d \vec{r} \left[ E_y(\vec{r},t) B_z(\vec{r},t) -  E_z(\vec{r},t) B_y(\vec{r},t)  \right]\nonumber \\
    &=-\frac{1}{32 \pi^4 c} \frac{d}{dt} \int d \vec{k} \left[ E_y(\vec{k},t) B_z(-\vec{k},t) -  E_z(\vec{k},t) B_y(-\vec{k},t)  \right].
\end{align}
We introduce Eqs. (\ref{3}), (\ref{4}) in Eq. (\ref{5}) and perform the integration over angles using
\begin{align}
\int d \Omega_{\vec{k}} \;k_{y,z}^2 k_x^p = \left\lbrace \begin{array}{l}
                                                          4 \pi k^{p+2}/(p+1)(p+3) , \; p \; \mbox{even}\\
                                                          0 , \; p \; \mbox{odd}
                                                          \end{array}   \right. . \nonumber
\end{align}
One obtains
\begin{align}\label{6}
F_x(t)=& - \frac{36 e^2}{\pi c^2 a^2} \frac{d}{dt} \sum_{\begin{array}{c}
       n, p=0\\
       n+p=\mbox{even}
       \end{array}}^{\infty} \frac{(-1)^n i^{n+p}}{n!p!(n+p+1)(n+p+3)} \int_0^{\infty} dr \int_0^{\infty} dr' \int_0^{\infty} dk\;\nonumber \\
       & \cdot k^{n+p} \left[ j_1(ka)\right]^2 \sin(kr) \sin(kr') w^p \left( t-\frac{r}{c}\right) \dot{w} \left( t-\frac{r'}{c}\right) w^n \left( t-\frac{r'}{c}\right),
\end{align}
and, after changing the summation indices $(n,p) \rightarrow (n, \kappa)$, $n+p=2\kappa$,
\begin{align}\label{7}
F_x(t)=&-\frac{36e^2}{\pi c^2 a^2} \frac{d}{dt} \sum_{\kappa=0}^{\infty} \sum_{n=0}^{2\kappa} \frac{(-1)^{n+\kappa}}{n! (2\kappa-n)! (2\kappa+1)(2\kappa+3)} \int_0^{\infty} dr \int_0^{\infty} dr' \int_0^{\infty} dk \nonumber \\
& \cdot k^{2\kappa} \left[ j_1(ka)\right]^2 \sin(kr) \sin(kr') w^{2\kappa -n} \left( t-\frac{r}{c}\right) \dot{w} \left( t-\frac{r'}{c}\right) w^n \left( t-\frac{r'}{c}\right).
\end{align}
To perform the integral over $k$, we use the method from \cite{fab}. We introduce the auxiliary integral 
\begin{align}\label{8}
I(a,b; r,r';\kappa) \equiv \int_0^{\infty} dk k^{2\kappa} j_1(ka)j_1(kb) \sin(kr) \sin(kr'),
\end{align}
and use \[ j_1(kx) = -\frac{1}{k^2} \frac{\partial}{\partial x} \frac{\sin(kx)}{x}.\] At the end, we will take $a=b$. In the following, to shorten the formulas, we will omit  $r,r'$ from the argument of $I$ .
We need to consider separately the cases $\kappa=0$, $\kappa=1$ and $\kappa \geq 2$.\\
\underline{\underline{Case $\kappa$=0}}
\begin{align}\label{9}
I(a,b;\kappa=0)= \frac{\partial}{\partial a} \frac{\partial}{\partial b} \frac{1}{ab} \int_0^{\infty} dk \frac{\sin(ka) \sin(kb) \sin(kr) \sin(kr')}{k^4}
\end{align}
For the integral $i=\int_0^{\infty}dk \frac{\sin(ka)\sin(kb)\sin(kr)\sin(kr')}{k^4}$, we use the result from the appendix, and for taking the derivatives required for the calculation of  $I(a,b;\kappa=0)$, we will use $\frac{d}{dx} sgn(x)=2\delta(x)$.\\
\underline{\underline{Case $\kappa$=1}}
\begin{align}\label{12}
I(a,b; \kappa=1) =\frac{\partial}{\partial a} \frac{\partial}{\partial b} \frac{1}{ab} \int_0^{\infty} dk \frac{\sin(ka) \sin(kb) \sin(kr) \sin(kr')}{k^2}.
\end{align}
We transform the products of trigonometric functions into sums, and  use Eq. 3.828 (2) from \cite{gra}
\begin{align}\label{13}
\int_0^{\infty} dx \frac{\sin(px) \sin(qx)}{x^2}= \frac{\pi}{2} \left[ p\; \theta(q-p)+ q\; \theta(p-q)\right].
\end{align}
After taking the derivatives, we will use the identities $\theta(x)- \theta(-x)=sign(x)$, where $sign$ is the signum function, and $x \cdot sign(x)=|x|$. After a little algebra, one obtains
\begin{align}\label{14}
I(a=b; \kappa=1)= \frac{\pi}{8}\left[ -\frac{|r-r'|}{a^4}+ \frac{r+r'}{a^4}+ \frac{4}{a^2}\delta(r-r')+2\frac{r-r'}{a^2} \delta'(r-r') \right],
\end{align}
where. again, we have written only the terms which will contribute to the integral over $r'$.\\
\underline{\underline{Case $\kappa \geq 2$}}
\begin{align}
I(a,b; \kappa \geq 2)= \frac{\partial}{\partial a} \frac{\partial}{\partial b} \frac{1}{ab} \int_0^{\infty} dk \;k^{2\kappa-4}\;\sin(ka) \sin(kb) \sin(kr) \sin(kr').
\end{align}
We transform the products of trigonometric functions into sums, and use $\int_0^{\infty} dx\; x^m \cos(kx)=\pi i^m \delta^{(m)}(k)$, where the upper index between parenthesis is the order of the derivative with respect to the argument. Keeping only the terms which will give contribution to the integral over $r'$, one obtains:
\begin{align}\label{16}
I(a=b; \kappa \geq 2)=& \frac{\pi}{8} (-1)^{\kappa} \left[ \frac{2}{a^4} \delta^{(2\kappa-4)} (r-r')-\frac{2}{a^2}\delta^{(2\kappa-2)}(r-r')-\frac{1}{a^4}\delta^{(2\kappa-4)}(2a+r-r') \right. \nonumber \\
& \left.-\frac{1}{a^2} \delta^{(2\kappa-2)}(2a+r-r')- \frac{1}{a^4} \delta^{(2\kappa-4)}(2a-r+r')-\frac{1}{a^2} \delta^{(2\kappa-2)}(2a-r+r') \right].
\end{align}
We can see from Eq. (\ref{16}) that $\int_0^{\infty} dr' I(a=b; r,r';\kappa)\cdot f(r') $ is a discontinuous function of variable $r$, because $\delta(2a-r+r')$ will give contribution only for $r>a$. So, if we try to evaluate the integral (\ref{8}) by expanding the spherical Bessel functions in Taylor series and integrating term by term, the result will not be accurate in the whole domain $r \in (0, \infty)$, because a discontinuous
function cannot be well approximated by a single power series.

With these results, we can return to Eq.(\ref{7}) and perform the integral over $r'$ and $r$, as explained in \cite{h}. We denote the results obtained for $\kappa=0$, $\kappa=1$, $\kappa \geq 2$ by $F_x^0$, $F_x^1$ and $F_x^2$ respectively. So, we have $F_x(t)=F^0_x(t)    + F^1_x(t)+ F^2_x(t).$ Compared to the calculations from \cite{h}, a supplementary difficulty appears in the evaluation of $F_x^1(t)$, because  the first two terms from Eq.(\ref{14}) are not Dirac delta functions. So, in the following, we present the detailed calculations only for  $F_x^1(t)$.

From Eqs. (\ref{7}), (\ref{14}), we obtain
\begin{align}\label{17}
F_x^1(t)=\frac{3e^2}{10c^2 a^2} \sum_{n=0}^{2} \frac{(-1)^n}{n!(2-n)!} \left[ -\frac{1}{a^4} f^1+\frac{1}{a^4} f^2 + \frac{2}{a^2} f^3\right],
\end{align}
where
\begin{align}\label{18}
f^1=\frac{d}{dt} \int_0^{\infty} dr \int_0^{\infty} dr'|r-r'| \;w^{2-n}\left(t-\frac{r}{c}\right) \dot{w} \left(t-\frac{r'}{c}\right) w^n \left(t-\frac{r'}{c}\right),
\end{align}
\begin{align}\label{19}
f^2=\frac{d}{dt} \int_0^{\infty} dr \int_0^{\infty} dr'(r+r')\; w^{2-n}\left(t-\frac{r}{c}\right) \dot{w} \left(t-\frac{r'}{c}\right) w^n \left(t-\frac{r'}{c}\right),
\end{align}
\begin{align}\label{20}
f^3=\frac{d}{dt} \int_0^{\infty}  w^{2}\left(t-\frac{r}{c}\right) \dot{w} \left(t-\frac{r}{c}\right).
\end{align}
In Eqs. (\ref{18}), (\ref{19}), we change the variables of integration $(r,r') \rightarrow (y,x)$, $t-r'/c=x$, $t-r/c=y$, and then we use the formula
\[ \frac{d}{dx}\int_0^xdy \int_0^x dz \;f(y,z)=  \int_0^x dz \; f(x,z)+ \int_0^x dy \; f(y,x),\]
which can be easily inferred from the Leibniz integral rule. One obtains
\begin{align}\label{21}
f^1=f^2=c^3 \left[ \int_{-\infty}^t dy (t-y)\;w^{2-n}(t)\dot{w}(y) w^n(y)+ \int_{-\infty}^t dy (t-y)\;w^{2-n}(y)\dot{w}(t) w^n(t)\right],
\end{align}
so, the contributions from $f^1$ and $f^2$ cancel each other in Eq. (\ref{17}). In Eq. (\ref{20}), we change the variable of integration $r\rightarrow x$, $t-r/c=x$ and we obtain
\begin{align}\label{22}
f^3=c w^2(t) \dot{w}(t).
\end{align}
From Eqs. (\ref{17}), (\ref{21}), (\ref{22}), one obtains $F_x^1(t)=0$. Analogously, we calculate  $F_x^0(t)$ and $F_x^2(t)$, and we obtain the final result
\begin{align}\label{23}
F_x(t)=& -\frac{6e^2}{ca^2} \dot{w}(t)+ \frac{9e^2}{2a^2} \sum_{\kappa=2}^{\infty} \sum_{p=0}^{\infty} \frac{a^{2p} \;2^{2p+2}\;(2p^2+7p+8)}{(2\kappa+1)(2\kappa+3)(2\kappa+1)!(2p+3)!(p+2)} \nonumber \\
& \cdot\frac{d^{2\kappa+2p+1}}{dr^{2\kappa+2p+1}} \left[w(t) - w\left(t-\frac{r}{c}\right) \right]^{2\kappa+1}_{|r=0}.
\end{align}
The first terms of the above equation are
\begin{align}\label{24}
F_x(t)=& -\frac{6e^2}{ca^2} \dot{w}(t)+ \frac{12e^2}{a^2} \frac{15(\beta^2-1) \tanh^{-1}\beta -2 \beta^5 -10 \beta^3 +15 \beta}{15\beta^2} \nonumber \\
& +\frac{34 \gamma^2}{5c^7} \left[ \frac{\gamma^2+2}{3} \dot{\ddot{w}}(t) (\dot{w}(t))^4 +\left(\gamma^4+\gamma^2+1\right) (\ddot{w}(t))^2 (\dot{w}(t))^3\right]+ {\cal O} (a^2),
\end{align}
where $\beta=\dot{w}(t)/c$, $\gamma=1/(1-\beta^2)$, and the summation over $\kappa$ has been performed as in \cite{h}, Eq. (28).  Using $ \tanh^{-1} \beta \cong \beta+\frac{\beta^3}{3} + \frac{\beta^5}{5}+ \frac{\beta^7}{7}+\dots,$ we obtain 
\begin{align}\label{25}
F_x(t)\cong & -\frac{6e^2}{ca^2} \dot{w}(t)+ \frac{144e^2}{7a^2} \beta^7 \nonumber \\
& +\frac{34 \gamma^2}{5c^7} \left[ \frac{\gamma^2+2}{3} \dot{\ddot{w}}(t) (\dot{w}(t))^4 +\left(\gamma^4+\gamma^2+1\right) (\ddot{w}(t))^2 (\dot{w}(t))^3\right]+ {\cal O} (a^2).
\end{align}

\section*{3. Conclusions}
We have calculated in this paper the electromagnetic self-force of a uniformly charged ball moving on a rectiliner trajectory, neglecting the Lorentz contraction. Our result (\ref{25}) contains a dominant term $ -\frac{6e^2}{ca^2} \dot{w}(t),$ which is directly proportional to the velocity and inversely  proportional to the square of the radius. It is different from the dominant term of the Sommerfeld-Page equation $-\frac{2e^2}{3c^2a}\ddot{w}(t)$, which is directly proportional to the acceleration and inversely proportional to the radius, and which is usually dropped because it is indistinguishable from the term $m \ddot{w}(t)$ in the equation of motion. This result supports the ideea that, even for small dimensions, the self-force is strongly dependent on the charge density.\\
\vspace{0.5cm}\\
{\bf Acknowledgement} The author is deeply grateful to Professor Vladimir Hnizdo for very useful e-mail correspondence.

\section*{Appendix}
\underline {\bf{The calculation of the integral $i=\int_0^{\infty}dk \frac{\sin(ka)\sin(kb)\sin(kr)\sin(kr')}{k^4}$ }}\\
\vspace{0.3cm}\\
This integral can be calculated integrating by parts in three steps. In the first step we use $\frac{1}{k^4}=\frac{d}{dk}\left(-\frac{1}{3k^3}\right)$, in the second step, $\frac{1}{k^3}= \frac{d}{dk}\left(-\frac{1}{2k^2}\right)$, and in the last step $\frac{1}{k^2}=-\frac{d}{dk}\frac{1}{k} $. One obtains
\begin{align}
i=\frac{\pi}{96}(i_1+i_2),
\end{align}
where
\begin{align}
i_1=& (a+b)^3[-sgn(a+b+r-r')-sgn(a+b-r+r') \nonumber\\
&+sgn(a+b+r+r')+sgn(a+b-r-r')]\nonumber\\
&+3(a+a)(r^2+r'^2)[-sgn(a+b+r-r')-sgn(a+b-r+r')\nonumber\\
&+sgn(a+b+r+r')+sgn(a+b-r-r')]\nonumber\\
&+6rr'[(a+b)sgn(a+a+r-r')+sgn(a+b-r+r')\nonumber\\
&+sgn(a+b+r+r')+sgn(a+b-r-r')]+(b \rightarrow -b),
\end{align}
\begin{align}
i_2=&3(a^2+b^2)(r-r')[-sgn(a+b+r-r')+sgn(a+b-r+r')\nonumber\\
&-sgn(a-b-r+r')+sgn(a-b+r-r')] \nonumber\\
&+6ab(r-r')[-sgn(a+b+r-r')+sgn(a+b-r+r') \nonumber\\
&+sgn(a-b-r+r')-sgn(a-b+r-r')] \nonumber\\
&+(r-r')^3[-sgn(a+b+r-r')+sgn(a+b-r+r')\nonumber\\
&-sgn(a-b-r+r')+sgn(a-b+r-r')]+ (r' \rightarrow -r').
\end{align}
Note that, for $a>b+r+r'$, one obtains eq. (3.746) from \cite{gra}.

\end{document}